\documentclass[twocolumn,showpacs,preprintnumbers,amsmath,amssymb]{revtex4}
\usepackage{graphicx}
\usepackage{dcolumn}
\usepackage{bm}

\begin{document}

\preprint{APS/123-QED}

\title{Guided atom laser: transverse mode quality and longitudinal momentum distribution}

\author{F. Vermersch$^{1,2}$}
\author{C. M. Fabre$^{1,2}$}
\author{P. Cheiney$^{1,2}$}
\author{G. L. Gattobogio$^{1,2}$}
\author{R. Mathevet$^{1,2}$}
\author{D. Gu\'ery-Odelin$^{1,2}$}

\affiliation{$^{1}$ Universit\'e de Toulouse ; UPS ; Laboratoire Collisions Agr\'egats R\'eactivit\'e, IRSAMC ; F-31062 Toulouse, France} 
\affiliation{$^{2}$ CNRS ; UMR 5589 ; F-31062 Toulouse, France}

\date{\today}

\begin{abstract}
We analyze the outcoupling of a matter wave into a guide by a time-dependent spilling of the atoms from an initially trapped Bose-Einstein condensate. This process yields intrinsically a breakdown of the adiabatic condition that triggers the outcoupling of the wave function. Our analysis of the time-dependent engineering and manipulation of condensates in momentum space in this context enables to work out the limits due to interactions in the mode quality of a guided atom laser. This study is consistent with recent experimental observations of low transverse excitations of guided atom lasers and suggests (i) an optimal strategy to realize such quasi-monomode guided atom lasers with, in addition, the lowest possible longitudinal velocity dispersion, or alternatively (ii) a strategy for engineering the atomic flux of the atom laser. 
\end{abstract}


\maketitle

Well-controlled coherent atom sources such as atom lasers are desirable for atom interferometry \cite{CSP09} and can be used as quantum probes of complex external potentials to study quantum transport phenomena \cite{qtransport,FCG11}. The first generation of atom lasers were obtained by outcoupling atoms from a trapped Bose-Einstein condensate (BEC) to free space \cite{FirstALgeneration}. Different aspects of these free falling atom laser have been extensively studied: the optimization and the characteristic of the output flux  \cite{RSH04,RMH05,RFH06,DDA10}, the longitudinal velocity dispersion \cite{JHH07} and the transverse mode quality \cite{RGL06,JDD08}.

 However, the de Broglie wavelength of a free falling atom lasers decreases rapidly because of the acceleration under gravity. To circumvent this problem a second generation of atom lasers involves atom outcoupling in an horizontal optical guide \cite{GRG06,CJK08,GCJ09,KWE10}. In this way, a mean de Broglie wavelength on the order of one micrometer can be achieved over macroscopic distances of a few mm.  

Similarly to free falling atom lasers, different outcoupling schemes have been demonstrated. In \cite{GRG06}, the trap is magnetic and atoms are outcoupled by radio-frequency spin flip. In \cite{CJK08}, atoms are extracted from a BEC held in a crossed dipole trap by applying an increasing magnetic field gradient along the guide direction. 
Alternatively, optical outcoupling was demonstrated and consists in lowering progressivelly the intensity of one of the beam of the crossed dipole trapped to favor the departure of atoms in the other beam of the trap \cite{GCJ09,KWE10,DHJ10}.

The reproducible preparation of a well-controlled propagating matter-wave is always a delicate task and is governed by the control of the extraction process. 
The mean number of modes occupied in the transverse direction can be easily infered from time-of-flight expansion for alkali atoms \cite{GRG06,CJK08,GCJ09} or by a detector based on a microchannel plate (MCP) for metastable helium atoms \cite{DHJ10}. Certain schemes prove the possibility to realize a quasi-monomode guided atom laser (GAL) while other schemes remain multimode with a small number of populated transverse modes \cite{CJK08,GCJ09,DHJ10}. Technical noise can affect dramatically the quality of the transverse mode of GAL. For instance, magnetic noise may lead to a limit in the control of rf outcoupling schemes \cite{BGB10}. However, even if one overcomes such experimental difficulties,  one may wonder, for a given outcoupling process, to which extent there exists a fundamental limit introduced by the outcouling mechanisms. A related question is to know what is the best strategy to realize an optimal outoucoupling for both longitudinal and transverse degrees of freedom. Indeed the longitudinal dispersion velocity governs the longitudinal coherence length of the atom laser. In all GAL experiments, there is a lack of information about this quantity and there is a need for methods allowing one to measure and engineer it \cite{GCG10}. Indeed narrow momentum distribution are required to obtain good fringe contrast in atom interferometry experiments \cite{CSP09}.

\begin{figure}[t!]
\centering
\includegraphics[width=7cm]{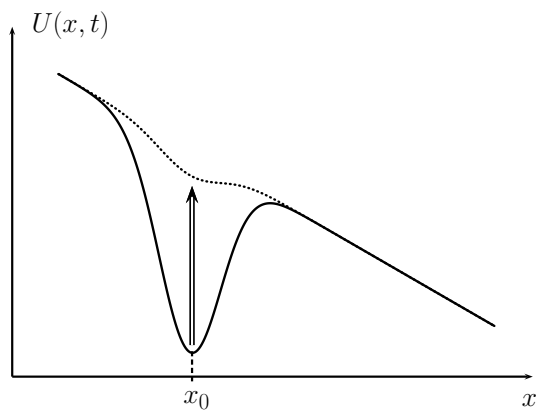}
\caption{Sketch of the optical outcoupling procedure. The intensity of the beam that ensures the longitudinal trapping is reduced progressively as a function of time which favors the spilling of the wave function outside the trap region.}
\label{Fig0}
\end{figure}

In this article, we focus on the study of the optical outcoupling scheme similar to the one implemented in \cite{GCJ09,KWE10,DHJ10}. Its principle is sketched in Fig.~\ref{Fig0}. Atoms are outcoupled from the trap by lowering its depth as a function of time. When the trap depth decreases, the strength of the trap also decreases and the minimum of the trapping potential is slightly displaced. At some point, even if the lowering of the trap depth is very slow, the adiabatic evolution breaks down since the timescale associated with the instantaneous trap diverges when the trap depth is reduced. This breakdown triggers the outcoupling of the wave function. Strictly speaking, this phenomenon rules out a priori any attempt based on a naive entropic point of view according to which the ground state wavefunction of the initially trapped atoms can be mapped on the ground state of the guide. To analyze this time-dependent problem, we mainly resort on numerical simulations. 

The paper is organized as follows. In Sec.~I, we discuss the impact of the breakdown of adiabaticity on the longitudinal momentum dispersion of the atomic wavefunction in a one-dimensional geometry and in the absence of interactions. In Sec.~II we take also into account the interactions within the mean field theory. In Sec.~III, we discuss the possibility of engineering the atomic flux using a simple model. In Sec.~IV, we use a two-dimensional geometry to account for the transverse modes in the guide and include the interactions. Our results explain the quality of the transverse mode obtained experimentally \cite{GCJ09} and provide a new strategy to reduce at best the longitudinal velocity dispersion requirements while preserving a very good transverse mode quality.


\section{1D outcoupling of an ideal Bose gas}

The potential experienced by the atoms is chosen in the form
\begin{equation}
U(x,t) = - m\gamma x- U_0 f(t) e^{-2x^2/w_0^2}.
\label{pot1D}
\end{equation}
The  function $f(t)$ characterizes the time dependence of the trap depth and tends to zero when time tends to infinity to ensure the outcoupling.
The other parameters are chosen to reproduce the typical experimental values: the mass $m$ is the one of rubidium-87 atom, the slope  (first term of the r.h.s. of Eq.~(\ref{pot1D})) is charaterized by an acceleration $\gamma$, and the angular frequency at the bottom of the trap is $\omega_0 = 2\sqrt{U_0}/w_0=2\pi \times 150$ s$^{-1}$ (depth $U_0/k_B=3.7$ $\mu$K and waist $w_0=40$ $\mu$m). The parameters are such that the time associated with the tunnel effect remains always extremely large compared to all other timescales of the problem.
The initial state of the wave function is here the ground state wave function associated with the harmonic oscillator potential obtained by the expansion of $U(x,t)$ to the second order in $x$ about the minimum $x=x_0$, $\partial_x U(x,0)|_{x_0}=0$ (see Fig.~\ref{Fig0}). 

\begin{figure}[t!]
\centering
\includegraphics[width=8cm]{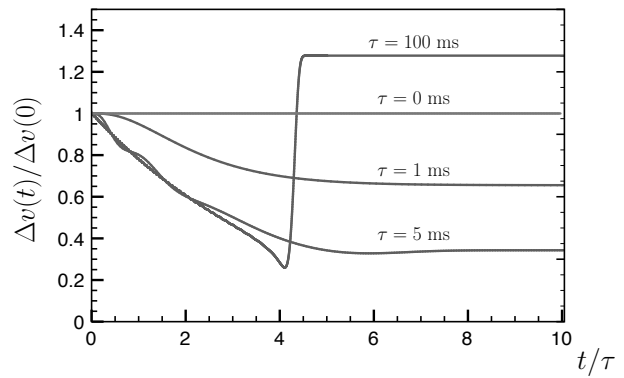}
\caption{Ratio between the longitudinal velocity dispersion of an ideal Bose-Einstein wavefunction as a function of time in the course of the outcoupling process and the initial velocity dispersion $\Delta v(0)= 0.58$ mm.s$^{-1}$ for different outcoupling characteristic times $\tau=0$ ms, $\tau=$ 1 ms, 5 ms and 100 ms. The time $t_O$ for which the potential does not have anymore a minimum  is $t_O\simeq$ 4$\tau$.}
\label{Fig2}
\end{figure}

We evolve the condensate wave function using a split-operator fast Fourier transform method \cite{FFS82}.
The evolution of the velocity dispersion in the course of the outcoupling processes is plotted in Fig.~\ref{Fig2} for an exponential time function $f(t)=e^{-t/\tau}$ and different characteristic times $\tau$. Such an exponential form has been used in \cite{CJK08}.
An abrupt switch off of the trap ($\tau=0$) enables to keep the velocity dispersion constant. Indeed, in this case, the whole wave function experiences only the linear potential slope and it is a well known result, used in time-of-flight experiments, that the velocity dispersion of a wave function is not affected when atoms are subjected to a constant acceleration field. When the characteristic time, $\tau$, is non-zero, the dispersion velocity starts to decrease. This is due to the beneficial effect of a quasi-adiabatic opening of the trap which increases the size of the cloud and decreases correspondingly the momentum dispersion since the ground state is a minimal state ($\Delta x_0 \Delta p_0 = \hbar/2$). The maximum gain in the reduction of the momentum dispersion is obtained at $t \sim t_O$ for $\tau \sim 5$ ms where $t_O$ is defined by
\begin{equation}
t_O = \tau \left( \ln \left(  \frac{2U_0}{m\gamma w_0} \right)-\frac{1}{2}  \right)\simeq 4\;\tau
\end{equation}
for the chosen exponential form of the time function $f(t)$ and corresponds to the time at which the potential does not have anymore a local minimum.

For longer time as used in experiments \cite{GCJ09,KWE10} (see Fig.~\ref{Fig2} for $\tau=100$ ms), three different phases can be distinguished. First, the velocity dispersion decreases as a result of the adiabatic opening of the trap. Second, a rapid increase occurs. This corresponds to the outcoupling stage. The large velocity dispersion originates from the fact that a part of the wave function is accelerated on the slope while the other part is still in the trap. This results in a stretching of the wave function in momentum space. The last stage corresponds to the propagation of the packet on the slope (constant momentum dispersion). The minimum value of the achievable velocity dispersion depends on the slope $\gamma$. In the limit of a vanishing slope, the lower bound is given by the Heisenberg limit $\hbar/2mw_0$. 

We conclude that, in the absence of interactions, an optimal strategy to minimize the longitudinal velocity dispersion, and referred in the following as the two stage strategy, consists in decompressing progressively the trap up to the point where outcoupling occurs (i.e. for $t\simeq t_O$) and then to switch it off abruptly so to freeze the velocity dispersion. In this way, it is possible to decrease the longitudinal velocity of the trapped atoms by a factor slightly larger than two  for $\gamma=0.2$ m.s$^{-2}$. This strategy does not depend on the specific shape of the function $f(t)$. It is worth mentionning that the adiabatic expansion stage can be speed up using shortcut methods as recently demonstrated \cite{MCR09,CRS10,SSV10,SSC11}.

\section{1D outcoupling of an interacting Bose-Einstein condensate}\label{section2}

\begin{figure}[t!]
\centering
\includegraphics[width=7.5cm]{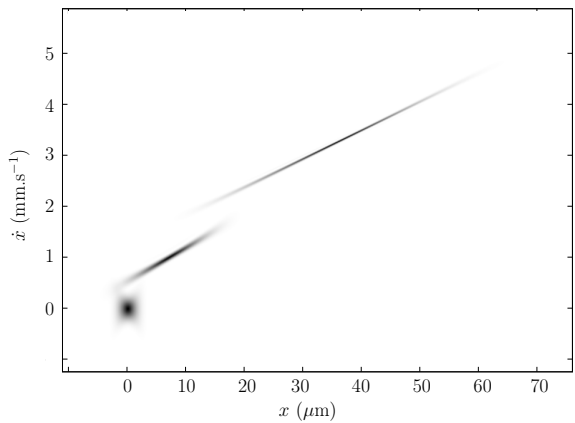}
\caption{Wigner function of the wave function at different times $t/\tau = 0$,  $4.5$ and $7$ for $f(t)=e^{-t/\tau}$ with $\tau=10$ ms. The wave function is evolved according to the 1D Gross Pitaevskii wave function with an initial condition corresponding to the Thomas Fermi regime ($Ng_1=53\,\hbar^{3/2}\omega_0^{1/2}m^{-1/2}$).}  
\label{Figwigner}
\end{figure}

The description of a condensate in the presence of repulsive interactions is made through an extra mean field term in the Schr\"odinger equation. The corresponding 1D Gross Pitaevskii (GP) reads
\begin{equation}
i\hbar \frac{\partial \psi}{\partial t}  = -\frac{\hbar^2}{2m}\frac{\partial^2 \psi}{\partial x^2}  + U(x,t)\psi  + g_1 |\psi |^2\psi, 
\label{1DGP}
\end{equation} 
where the wave function $\psi$ is normalized to the total number of particles.
At $t=0$, the BEC wave function is taken to be a stationary solution of the time independent GP equation. It is found by evolving the time-dependent GP equation in imaginary time $t \rightarrow i t$ in the potential (\ref{pot1D}) with $\gamma=0$ (no slope) but taking into account the displacement of the minimum $x_0$ induced by the slope. This trick facilitates the numerical convergence towards the stationary state. 
We have chosen the strength parameter, $Ng_1=Ng_1^{\rm ex}=53\,\hbar^{3/2}\omega_0^{1/2}m^{-1/2}$, so that the condensate is well in the Thomas Fermi regime and the density profile is close to an inverted parabola: 
\begin{equation}
n_1(x,t=0)=\frac{3N}{4R_1^0}\left(1-\frac{x^2}{(R_1^0)^2} \right)
\label{density1D}
\end{equation} 
where the Thomas Fermi radius reads $R_1^0=(3Ng_1/2m\omega_0^2)^{1/3} \simeq$ 3.8 $\mu$m with our parameters \cite{footnote0}.
With these parameters close to those used in experiments \cite{GCJ09,KWE10}, $\Delta v(0)= 0.25$ mm.s$^{-1}$. The interaction energy per particle reads $E^{\rm int}_1(R_1^0)=2\mu_1^0/5=3Ng_1/10 R_1^0$ where $\mu_1^0$ is the chemical potential. 

\begin{figure}[t!]
\centering
\includegraphics[width=7.5cm]{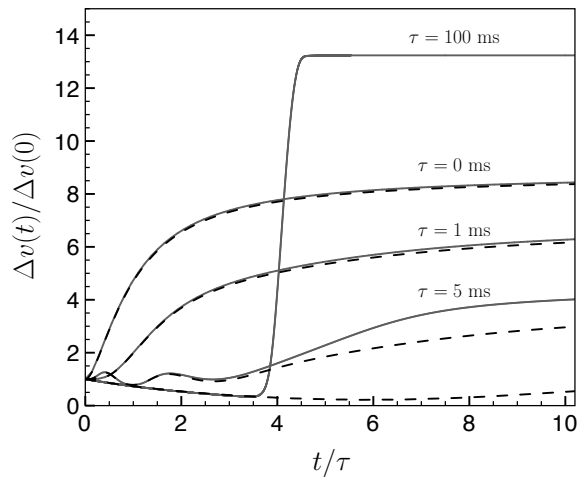}
\caption{Ratio between the longitudinal velocity dispersion of a 1D wave packet and  the initial velocity dispersion $\Delta v(0)= 0.25$ mm.s$^{-1}$ as a function of time in the course of the outcoupling process  for (i) a sudden switch off of the trap (represented in units $\tau=1$ ms), and (ii) for different outcoupling characteristic time:  $\tau=$1 ms, $\tau=$5 ms, $\tau=$100 ms. Gray solid lines: numerical simulation. Dashed lines: results from the analytical model in momentum space for the opening of the trap detailed in Appendix A. The breakdown of the good agreement between simulation and the model occurs when the spilling of atoms starts.
The wave packet describes the wave function of a condensate initially in the Thomas Fermi regime with an interaction parameter $Ng_1=53\,\hbar^{3/2}\omega_0^{1/2}m^{-1/2}$.}
\label{Fig3}
\end{figure}

Figure \ref{Figwigner} shows the Wigner function in a phase space representation of the wave function at three different times: in the trap before the outcoupling, just after the outcoupling procedure, and in the course of the expansion. A space-velocity correlation develops as a function of time. As a result, the analysis in position space of the interaction of an atom laser with an extra potential provides directly the behavior for the different class of velocity present in the wave packet. This feature is interesting in view of using atom lasers as quantum probes \cite{FCG11}.

The evolution of the momentum dispersion of the wave function as a function of time is plotted in Fig.~\ref{Fig3} for different characteristic times $\tau$. 
The momentum dispersion increases at short time $t<\tau$ (before any outcoupling) for low values of the characteristic time $\tau=0,\,1$ ms but decreases for longer characteristic times $\tau=100$ ms. 
This means that two contributions with opposite effect are involved in the dynamics of the momentum dispersion. At an intermediate characteristic time $\tau=5$ ms, this trade-off yields 
a transient oscillatory behavior (see Fig.~\ref{Fig3}). The increase is well captured by the standard analysis based on scaling parameters of a Thomas Fermi cloud. It is due to the conversion of interaction energy into
kinetic energy. The decrease is attributed to the adiabatic expansion and corresponds to the decrease of the proper kinetic energy. The dynamics of the trade-off in between those two contributions is discussed in Appendix A using an extension of the standard scaling ansatz method. The corresponding predictions based on an harmonic approximation of the bottom of the trap are shown as dashed lines in Fig.~\ref{Fig3}. They capture very well the evolution of the momentum dispersion in all range of values for the characteristic time $\tau$,
and are valid during the opening of the trap and before the outcoupling of atoms.

 For short characteristic times ($\tau <$1 ms), we obtain a poor gain (on the order of 10 \%) in the asymptotic longitudinal dispersion velocity compared to its value for an abrupt switch off of the trap. For $\tau \simeq 5$ ms, a gain up to 50 \% can be obtained as a result of the beneficial effect of the adiabatic opening of the trap. 
As previously, the two stage strategy is still efficient to get the smallest possible longitudinal velocity dispersion. However, such a strategy prohibits to engineer the atomic flux (see next section).
  
When the strength of the repulsive interactions is increased, the initial size of the condensate increases ($R \propto g_1^{1/3}$ in the Thomas Fermi regime) and the momentum dispersion decreases compared to its value in the absence of interactions. However and in contrast with the ideal Bose gas case, the presence of repulsive interactions yields to an increase of the asymptotic velocity dispersion $\Delta v(\infty)$ compared to its initial value $\Delta v(0)$ in all schemes, i.e. whatever is the characteristic time $\tau$. This is due to both the transfer of the interaction energy into kinetic energy in the course of the expansion of the wave packet (dominant process at low $\tau$ values) and the stretching effect of the wavefunction in the course of its outcoupling (dominant process at large $\tau$ values). For instance, assuming $\tau = 5$ ms, we find $\Delta v (t\rightarrow \infty ; g_1 = g_1^{\rm ex}) \simeq 5.5 \Delta v (t\rightarrow \infty ; g_1 = 0)$. 
Interestingly, the timescale to reach the asymptotic value turns out to be quite robust with the interaction strength. Indeed, the timescale deduced from the ratio between the Thomas Fermi radius and the sound velocity turns out to be independent of $g_1$ and on the order of $\omega_0^{-1}$ (this latter result is even valid in all space dimensions and explains why the surface modes of a dilute condensate are independent of the interaction strength \cite{DGP99}).

\begin{figure}[t!]
\centering
\includegraphics[width=7.5cm]{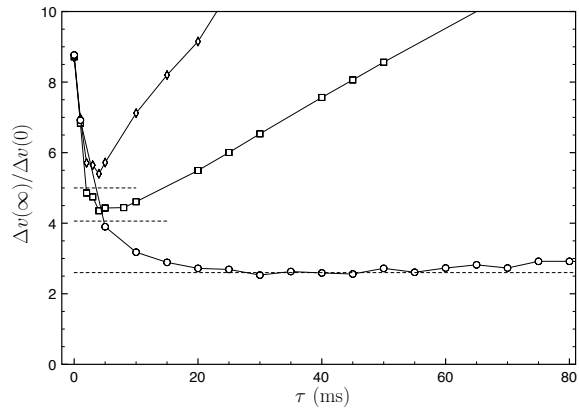}
\caption{Asymptotic longitudinal velocity dispersion ratio $\Delta v(\infty)/\Delta v(0)$ as a function of the characteristic time $\tau$ used for the outcoupling process with an exponential time function $f(t)=\exp(-t/\tau)$: $\gamma=1$ m.s$^{-2}$ (open diamonds),  $\gamma=0.2$ m.s$^{-2}$ (open squares) and $\gamma=0.01$ m.s$^{-2}$ (open circles). Same initial conditions as in Fig.~\ref{Fig3}. The dotted lines correspond to the lower bound estimated with Eq.~(\ref{vlimit}).}
\label{figureg}
\end{figure}

The slope (parameter $\gamma$) limits ultimately the minimum achievable asymptotic velocity dispersion as illustrated in Fig.~\ref{figureg}.
The value $\Delta v(\infty)$ obtained for $\tau$ very small ($<3$ ms) is always larger than the initial dispersion velocity as discussed already (see Fig.~\ref{Fig3}). 
The shape of those curves corresponding to different values of $\gamma$ is well understood: the initial decrease is due to a beneficial adiabatic opening and the increase to the stretching of the wave function and thus to the slope. In the absence of a slope, only the decrease is observed. 

In order to estimate the lowest achievable asymptotic dispersion we proceed in the following way. For each value of the slope coefficient $\gamma$, we calculate with a gaussian ansatz the maximum radius size $\tilde{R}$ of the wavefunction for the lowest depth compatible with a stationary solution. By decreasing further the trap depth, the absence of a stationary solution corresponds to the onset of the outcoupling process, and thus to the breaking of the adiabaticity. The corresponding asymptotic velocity dispersion is then obtained by energy conservation at this threshold value for the Thomas Fermi radius: 
\begin{equation}
(\Delta v)_{\rm min} = \left[ \frac{2}{m}  \left(  K_1(\tilde{R})+E_1^{\rm int}(\tilde{R}) \right) \right]^{1/2},
\label{vlimit}
\end{equation}
where $K_1(R)$ is the proper kinetic energy per particle of the BEC in the Thomas Fermi regime (see Appendix) and $E^{\rm int}_1(R)$ is the corresponding interaction energy per particle. We have represented these limits as dotted lines in Fig.~\ref{figureg}. These minima turn out to be robust against the specific shape of the time function $f(t)$. They give a good estimate of the minimum achievable asymptotic velocity dispersion. For a residual slope parameter of $\gamma=0.01$ m.s$^{-2}$, the achievable velocity dispersion is ten times lower than the recoil velocity. Such a technique has been used in \cite{FCG11} to prepare a BEC for probing an optical lattice.

\section{Flux engineering}

\begin{figure}[t!]
\centering
\includegraphics[width=7.5cm]{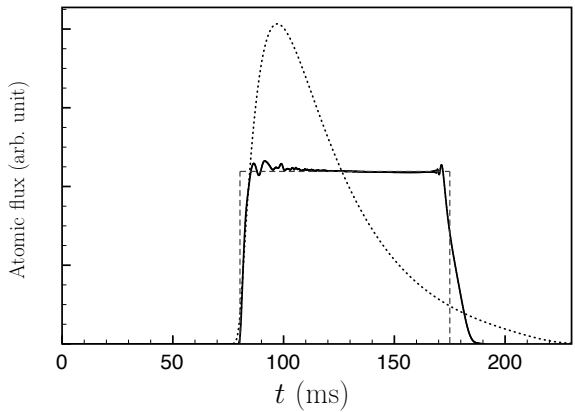}
\caption{Output flux as a function of time for a time evolution function (i) $f(t)=\exp (-t/\tau)$ (dotted line) and (ii) $f(t)$ (solid line) deduced from Eq.~(\ref{vdepthproper}) with a desired square flux $J_0$ (dashed line). Same initial parameters as in Fig.~\ref{Fig3}.}
\label{sshaping}
\end{figure}

Another important question from the experimental point of view is the possibility of controlling the output flux by a proper choice of the time function $f(t)$. Let us consider the examples depicted in Fig.~\ref{sshaping}. With the same initial parameters as those of Fig.~\ref{Fig3}, the decrease of the trap depth is sufficiently small to allow outcoupling above a time  $t_d \simeq$80 ms. With an exponential form $f(t)=\exp (-t/\tau)$ for the time evolution of the trap depth, the output flux $J(t)=-{\rm d}N/{\rm d}t$ (where $N(t)$ is the number of trapped atoms) obtained numerically has the shape of the dotted curve. 

In the following, we model the potential as a linear slope plus a harmonic potential whose angular frequency is kept constant. This simplification allows for an analytic solution and does not affect the generality of our approach and its conclusion. We show how a quasi-equilibrium assumption helps to engineer the outcoupled atomic flux. 

The outcoupling is provided here by reducing the depth of the harmonic potential but not its strength. 
For $t<t_d$, the chemical potential $\mu^0$ is smaller than the trap depth $V_{\rm depth}(t<t_d)$ and no atoms are outcoupled. For $t>t_d$, we assume that the timescale over which the trap is decreased is sufficiently low to ensure the locking of the chemical potential to the instantaneous trap depth: $\mu(t)=V_{\rm depth}(t>t_d)$. Assuming that the relation between the chemical potential and the number of atoms is still valid in this quasi-equilibrium regime, we can infer the time evolution of the number of atoms $N(t)$ and thus of the atomic flux $J(t)$.

Conversely, if we impose a desired flux function $J_0(t)$, we can provide the corresponding time-evolution of the trap depth. In 1D, this relation is given by 
\begin{equation}
V_{\rm depth}(t)=\mu^0 f(t)=\mu^0 \left[ \left( 1-\frac{1}{N(0)}\int_{t_d}^t J_0(t) {\rm d}t \right) \right]^{2/3},
\label{vdepthproper}
\end{equation}
for $t>t_d$. Figure~\ref{sshaping} provides an example of flux engineering. $J_0$ is taken with a square (dashed line) form in order to impose a nearly constant flux for a large amount of time. The output flux (solid line) obtained from the numerical simulation performed with $V_{\rm depth}(t)$ given by Eq.~(\ref{vdepthproper}) gives a satisfying result compared to the desired flux. 

The function $f(t)$ has also a strong influence on the asymptotic velocity dispersion once all atoms are outcoupled. As an example, for the square prescription $J_0$, the constraint on the atomic flux yields an increase by a factor five of the asymptotic velocity dispersion compared to the optimal two stage strategy.



\section{2D outcoupling of an interacting Bose-Einstein condensate}

The 1D models developed in the Sec. I and II give the optimal strategy for minimizing the longitudinal velocity dispersion. In this section we consider a 2D model which therefore accounts also for the transverse degrees of freedom. Our starting point is the 2D Gross-Pitaevskii 
\begin{equation}
i\hbar \frac{\partial \psi}{\partial t}  = -\frac{\hbar^2}{2m}\left(\frac{\partial^2 \psi}{\partial x^2}+\frac{\partial^2 \psi}{\partial y^2}\right)  + U(x,y,t)\psi  + g_2 |\psi |^2\psi, 
\label{1DGP}
\end{equation} 
where the wave function $\psi$ is normalized to the total number of particles. A first set of simulations has been performed with the 2D trapping potential 
\begin{equation}
U_{\rm I}(x,y,t) =  - m\gamma x- U_0 f(t) e^{-2x^2/w_0^2} + \frac{1}{2}m\omega_y^2y^2,
\end{equation}
where the guide is accounted for by a transverse harmonic potential of angular frequency $\omega_y$ and $f(t)=e^{-t/\tau}$. 
In practice, the Gaussian potential associated with the trap can be realized using a cylindrical red detuned off resonance Gaussian beam with a large waist along the direction perpendicular to the propagation direction $x$.
As expected, the results that we have obtained for the longitudinal degrees of freedom are very close to those detailed in the previous section. Here, a sudden switch off is realized by setting $U_0$ to zero while keeping the guide confinement unchanged. The asymptotic value of the longitudinal momentum dispersion for a sudden switch off can be deduced from Eq.~(\ref{app7}) of appendix A.

\begin{figure}[htbp]
\centering
\includegraphics[width=8cm]{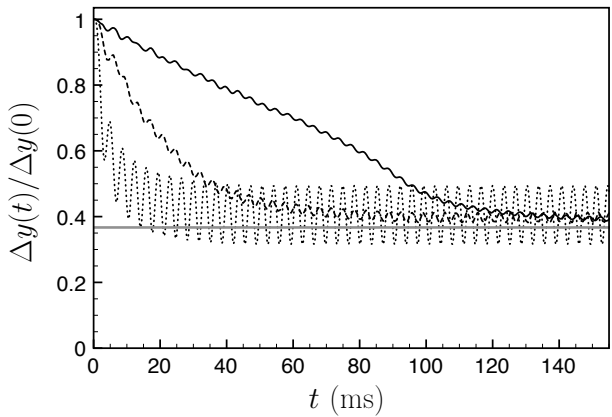}
\caption{Transverse quadratic size as a function of time in the course of the outcoupling process for different outcoupling characteristic time $\tau=$ 1 ms (dotted line), 5 ms (dashed line) and 20 ms (solid line). The gray straight line corresponds to the quadratic size of the harmonic oscillator associated with the transverse confinement of the guide.}
\label{Fig5}
\end{figure}

\begin{figure}[htbp]
\centering
\includegraphics[width=8cm]{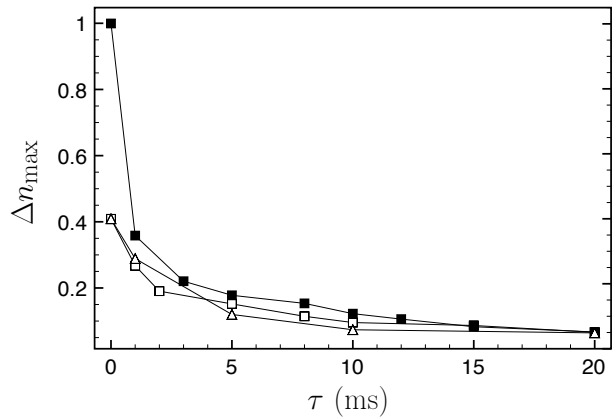}
\caption{Upper bound of the contamination of the transverse degree resulting from the outcoupling as a function of the outcoupling characteristic time $\tau$. Calculations are performed for a strength parameter of $Ng_2=267\,\hbar^2/m$ and potentials $U_{\rm I}$ (open squares and diamonds) and $U_{\rm II}$ (black squares). An exponential decay $f(t)=\exp(-t/\tau)$ is chosen for open and black squares while a linear decay $f(t\leq \tau)=1-t/\tau$ has been chosen for data represented by open diamonds. The potential $U_{\rm II}$ which couples strongly the longitudinal and transverse degrees of freedom turns out to give similar result as $U_{\rm I}$ when the outcoupling time is sufficiently long ($\tau>10$ ms for our parameters).}
\label{Fig6}
\end{figure}

Figure \ref{Fig5} gives examples of the variation of the transverse dispersion $\Delta y$ as a function of time in the course and after the outcoupling for
different characteristic times $\tau=1$, $5$ and $20$ ms. The strength of the interactions, $Ng_2=267\,\hbar^2/m$, has been chosen so to have the same Thomas Fermi radius as for the 1D case with $Ng_1=53\,\hbar^{3/2}\omega_0^{1/2}m^{-1/2}$. As expected, a too fast outcoupling is accompanied with large transverse oscillations. 
The longer the outcoupling time $\tau$, the lower the transverse excitation.  
A more quantitative analysis is summarized in Fig.~\ref{Fig6}. We have plotted an effective maximum quantum number $ \Delta n_{\rm max} $ associated with the contamination of the transverse degrees of freedom using the formula
\begin{equation}
(\Delta y)_{\rm max}^2 = \frac{\hbar}{m\omega_y} \left( \Delta n_{\rm max} + \frac{1}{2} \right)
\end{equation}
where $(\Delta y)_{\rm max}^2$ corresponds to the asymptotic maximum amplitude of the transverse dispersion. 
$ \Delta n_{\rm max} $ thus measures the contamination of the transverse excited states due to the outcoupling process.
We observe a situation in which the first excited state population remains below 10 \% for $\tau > 10$ ms for our parameters. Note, however, that the amount of
contamination for a given $\tau$ increases slightly with the strength $g_2$ of the interactions.

In the limit of a sudden release of the condensate into the waveguide, the expansion can be accounted by a scaling law approach and the modal decomposition can be worked out \cite{PlS02,StZ02}. One finds analytically and numerically that a frequency mismatch between the initial trap and the guide angular frequency triggers transverse oscillations. The larger the guide frequency the larger the transverse amplitude. However, if one performs a progressive spilling of the atoms into the guide, these oscillations can be dramatically reduced similarly as in Fig.~\ref{Fig5}. If the guide frequency is larger than the trap frequency, a relatively low characteristic time, $\tau$, is required. In the opposite limit, a relatively large $\tau$ is necessary to damp out the transverse oscillations. 

The potential $U_{\rm I}$ minimizes by construction the coupling between the transverse and longitudinal degrees of freedom.
We have also investigated the same problem with the potential 
\begin{equation}
U_{\rm II}(x,y,t) = - m\gamma x- U_0 e^{-t/\tau} e^{-2(x^2+y^2)/w_0^2} + \frac{1}{2}m\omega_y^2y^2.
\end{equation}
closer to the one used in the experiments of Refs.~\cite{GCJ09,KWE10}. For both potentials, the upper value $ \Delta n_{\rm max}$ remains small ($<0.1$) for $\tau>10$ ms. This is in agreement with the qualitative estimate ( $\Delta n^{\rm int}<0.1$) of the contribution to $\Delta n$ of the interactions proposed in Ref.~\cite{GCJ09}. In addition, the specific shape $f(t)$ of the decrease of the trapping potential does not play a crucial role on the transverse state populations for a sufficiently large characteristic time $\tau$ as illustrated in Fig.~\ref{Fig6} where the results for a linear and an exponential form of $f(t)$ are compared.

The long outcoupling times chosen in the experiments \cite{GCJ09,KWE10} are optimal to reduce the contamination of the transverse excited states and explain the possibility of reaching the quasi monomode regime for guided atom lasers. However, they are not optimal from the longitudinal dispersion point of view.

As in Sec.~\ref{section2}, we find numerically that the optimal and robust strategy is still the two stage strategy which consists in an adiabatic opening of the trap (just before the time for which atoms can spill into the guide) followed by an abrupt switch off of the trapping potential. Indeed, the value obtained for $\Delta n_{\rm max}$ turns out to be robust against the sudden switch off of the trap.  As a complementary procedure, one could envision to compensate the local slope by an external potential to enable a more efficient adiabatic opening. These features are quite general and expected to be still valid in 3D \cite{footnote}. However, as in 1D, it is not possible to engineer the atomic flux and to benefit from the lowest achievable asymptotic velocity dispersion.

In conclusion, we have presented a detailed study of the spilling outcoupling mechanism to produce a guided atom laser in both position and momentum space. The main features and limits in velocity dispersion have been discussed and can be captured by simple models. The detail study of the control in momentum space of an interacting condensate reveals the importance of the  two contributions to the kinetic energy: the proper kinetic energy of the condensate and the one associated with the time-dependent transformation of the potential experienced by the condensate. Our study enables to optimize either the atomic flux or the momentum space of a quantum probe realized by spilling atoms from a BEC. In addition, it validates the entropic analysis performed in \cite{GCJ09} since non-adiabatic and interactions effects have been shown to play a minor role for the chosen experimental protocol.


\section*{ACKNOWLEDGEMENTS}
We are grateful to T. Lahaye  for useful comments. We acknowledge financial support from the R\'{e}gion Midi-Pyr\'{e}n\'{e}es, the C.N.R.S., the Agence Nationale de la Recherche (ANR-09-BLAN-0134-01) and Institut Universitaire de France.

\appendix

\section{Momentum dispersion of a non-ideal condensate in a time-dependent harmonic trap}

The time evolution of the shape of a condensate in the Thomas Fermi regime and under a time-variation of the trap frequencies can be accounted exactly for by a set of coupled equations involving scaling factors \cite{KSS96,CaD97}. In this appendix, we show that this scaling approach captures the size evolution of the condensate but fails to take properly into account the evolution of the momentum dispersion. Indeed, Thomas Fermi approximation yields a divergency of the kinetic energy at the boundary of the trapped condensate. 

The explicit calculation of the kinetic energy where the boundary is carefully treated through a suitable expansion is detailed in Ref.~\cite{DPS96} for a 3D harmonically trapped BEC in the Thomas Fermi regime. 

We give here explicitly an approach which predicts correctly the mean evolution of the momentum $\langle p^2 \rangle$ of a harmonically trapped BEC when the strength of the confinement is modified as a function of time. The initial angular frequency is denoted $\omega (0)=\omega_0$ and is assumed to be the same in all directions. The calculation is performed at $d=1,2,3$ dimensions. 

In order to work out the time evolution of the velocity dispersion we have to calculate the time-dependent mean kinetic energy per particle along a given axis
\begin{equation}
\frac{\langle p_x^2 \rangle(t)}{2m} = \frac{\hbar^2}{2m} \frac{1}{N}  \int_{-\infty}^\infty \left| \frac{ \partial \psi (\vec{r},t) }{ \partial x }  \right|^2 {\rm d} x.
\label{kinenergy}
\end{equation}
Let us introduce $n_0(\vec{r}\,)$ the \emph{exact} initial density profile before modifying the trap strength. This profile is essentially the one of an inverted parabola except in the vicinity of the boundaries \cite{DPS96}. Its initial Thomas Fermi radius $R^0_d$ is related to the chemical potential $\mu^0_d$ by $R^0_d=(2\mu_d^0/m\omega_0^2)^{1/2}$. The kinetic energy per particle for a BEC at equilibrium in the harmonic trap is a function of $R^0_d$: 
\begin{equation}
K_d[R^0_d]= \frac{A_d\hbar^2}{m(R^0_d)^2} {\rm ln} \left[  (R^0_d)^{(2-d)/4}\left( \frac{mg_dN}{\hbar^2}\right)^{1/4}B_d \right],
\end{equation}
where the dimensionless coefficients $A_d$, $B_d$  ($A_1=0.50$, $A_2=1.17$, $A_3=2.50$, $B_1=1.52$, $B_2=1.45$, $B_3=0.80$) have been determined numerically following the procedure detailed in Ref.~\cite{DPS96} for the 3D case. This method consists in matching the asymptotic form of the wave function in the interior of the cloud and in the boundary region and has been generalized here  to 1D and 2D case.

\begin{figure}
\centering
\includegraphics[width=8cm]{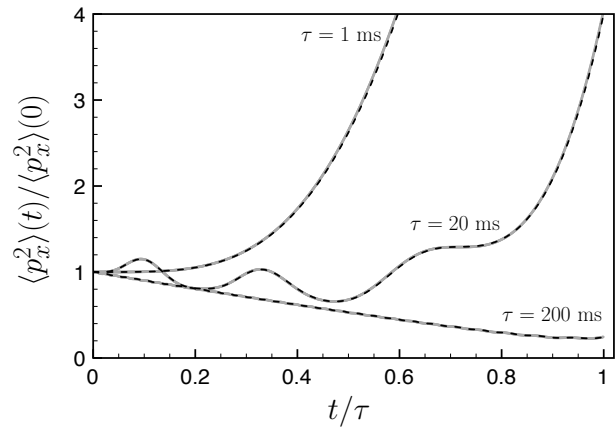}
\caption{Time evolution of the kinetic energy of an interacting Bose-Einstein condensate (same initial conditions as in Fig.~\ref{Fig3}) held in an harmonic trap of initial angular frequency $\omega_0=2\pi \times 150$ Hz when the angular frequency is weakeaned as a function of time according to $\omega(t)=4\omega_0(1-t/\tau)/5+\omega_0/5$: Numerical simulation (solid gray line), scaling law approach (see Eq.~(\ref{kinenergy2})) (dashed line). A very good agreement is observed for various values of $\tau=1, \,20,\, 200$ ms.}
\label{appendix}
\end{figure}

The scaling method suggests to approximate the time-dependent condensate wave function by
\begin{equation}
\psi_d (\vec{r},t) = \left[\frac{1}{b^d(t)}n_0\left(\frac{\vec{r}}{b^(t)} \right) \right]^{1/2}\exp \left( \frac{im\dot{b}}{2b\hbar} r^2\right),  
\label{ansatz}
\end{equation}
where the scaling parameter $b(t)$ obeys an Ermakov-like equation
\begin{equation}
\ddot{b} + \omega^2(t) b = \frac{\omega_0^2}{b^{d+1}},
\label{evolutiondeb}
\end{equation}
with the initial conditions $b(0)=1$ and $\dot{b}(0)=0$. 

To carry out the calculation of the kinetic energy, we first consider the 1D situation. The kinetic energy term (\ref{kinenergy}) involves the space derivative of the wave function.
The ansatz (\ref{ansatz}) has two position dependent terms originating from the modulus $\rho_1(x,t)$ and phase $S_1(x,t)$ of the wave function. Its derivative thus reads 
\begin{equation}
\frac{\partial \psi_1(x,t)}{\partial x} = \left( \frac{\partial \rho_1(x,t)}{\partial x}  + i\rho_1(x,t)\frac{\partial S_1(x,t)}{\partial x} \right) e^{iS_1(x,t)}.
\end{equation}
The modulus square $|\partial \psi_1(x,t)/\partial x|^2$ involves two square terms that are calculated with different approximation schemes. The contribution to the kinetic energy of the square term $|\partial \rho_1/\partial x|^2$ requires to include the contribution of the wings to cancel out artificial divergences, and is equal to $K_1[b(t)R_0]$. The contribution to the kinetic energy of the square term $|\rho_1\partial S_1/\partial x|^2$ is finite and can be carried out within the Thomas Fermi approximation yielding a term that scales as $\dot{b}^2$. Such a procedure can be generalized to higher dimenisions. We finally get
\begin{equation}
\frac{\langle p_x^2 \rangle  (t)}{2m}=\frac{K_d[b(t)R_d^0]}{d}+ \frac{C_d\mu^0_d}{\omega_{0}^2} \left( \frac{{\rm d} b(t) }{ {\rm d} t}\right)^2,
\label{kinenergy2}
\end{equation}
where $C_1=1/5$, $C_2=1/6$ and $C_3=1/7$, and the relation between the chemical potential and the interaction strength is $\mu^0_1=3Ng_1/4R_1^0$, $\mu^0_2=2Ng_2/\pi (R_2^0)^2$, $\mu^0_3=15 N g_3 /8\pi(R_3^0)^3$.

Equation (\ref{kinenergy2}) shows the two contributions to the kinetic energy. Depending on the characteristic time over which the dimensionless scaling parameter $b$ evolves, the dominant contribution to the momentum dispersion changes: (i) in the adiabatic limit, the time derivative tends to zero and the first terms accounts for the scaled variation of the kinetic energy (case $\tau=100$ ms in Fig.~\ref{appendix}); (ii) for a rapid change of the confinement such as in an abrupt switch off the second term dominates (case $\tau=1$ ms in Fig.~\ref{appendix}). For intermediate timescale, the two contributions yields oscillations of the momentum dispersion (case $\tau=20$ ms in Fig.~\ref{appendix}). We find a remarkable agreement between the numerical integration of the Gross-Pitaevskii equation and formula (\ref{kinenergy2}). 

Note that the second term of the left-hand side of Eq.~(\ref{kinenergy2}) that accounts for the transfer of the interaction energy into kinetic energy in the course of the expansion can be directly obtained by combining the moment method (via the Ehrenfest theorem) with the scaling parameters \cite{Gue02,PGS03,ImG10}.

For an abrupt switch off ($\omega(t>0)=0$), we can extract from Eqs.~(\ref{evolutiondeb}) and (\ref{kinenergy2}), the complete dynamics of the evolution of the momentum dispersion. In particular, the asymptotic increase value of the momentum dispersion reads 
\begin{equation}
\frac{\langle p_x^2\rangle (\infty)}{2m} \simeq \frac{2C_d\mu_d^0}{d}=\frac{E_d^{\rm int}(t=0)}{d}.
\label{app7}
\end{equation}


\begin{thebibliography}{99}
\bibitem{footnote0} It corresponds to the Thomas Fermi radius of a 3D BEC of 20000 atoms in an isotropic harmonic trap of angular frequency $\omega_0=2\pi \times 150$ Hz.
\bibitem{footnote} The important adiabatic decompression stage can be speed up in principle for a 2D condensate initially in the Thomas-Fermi regime using adiabaticity shortcut methods \cite{MCR09}.
\end{thebibliography}
\end{document}